\documentclass[twocolumn,aps,floats,floatfix,nofootinbib,prd,superscriptaddress,tightenlines]{revtex4}

\def\pslash{p\!\!\!\slash}
\def\nslash{n\!\!\!\slash}

\begin{document}

\title{Spin dependent masses and $Sim(2)$ symmetry}
\author{JiJi Fan, Walter D. Goldberger, and Witold Skiba}
\affiliation{Department of Physics, Yale University, New Haven, CT 06520}

\begin{abstract}
Recently, Cohen and Glashow pointed out that all known experimental tests of relativistic kinematics are consistent with invariance of physics under the four-parameter subgroup $Sim(2)$ of the Lorentz group.   The massive one-particle irreducible representations of  $ISim(2)$,  that is $Sim(2)$ times spacetime translations, are all one-dimensional, labeled by spin along a preferred axis. Consequently particle theories based on this symmetry can accomodate lepton number conserving masses for left-handed neutrinos without the need to introduce sterile states.  The same property of massive particle representations, however, also leads to the possibility that particle masses may be split within the diffferent spins of a representation of the ordinary Poincare group.
 In this article we investigate  the low-energy structure of theories with spin dependent masses and comment on the bounds
on such effects.
\end{abstract}

\maketitle
\section{Introduction}

It is usually assumed that spacetime symmetries can only break spontaneously through the emergence of some non-zero order parameter, for instance a classical background tensor field, that carries spacetime quantum numbers.   This order parameter then controls the magnitude of the symmmetry breaking effects.  It also serves as a `spurion' for constructing local operators in the Lagrangian that mediate the effects of the broken symmetry while at the same time preserving the group $H$ of spacetime tranfromations that remain unbroken.

Recently, Cohen and Glashow~\cite{CG} pointed out that for some choices of $H$ the effects of symmetry breaking cannot be described in this way.   It is possible, for instance, for the $SO(3,1)$ Lorentz invariance of physics to break down to the subgroups $Hom(2)$ or $Sim(2)$.   These subgroups do not admit invariant tensors which could act as spurions.   Consequently, the breaking of $SO(3,1)$ to either $Hom(2)$ or $Sim(2)$ cannot be described in terms of local, symmetry breaking operators.   This situation was referred to as ``very special relativity" (VSR) by the authors of~\cite{CG}.

There are many theories that might exhibit minute departures from Lorentz symmetry at low energies,
see for example Refs.~\cite{ Lv1, Lv2, Lv3, Lv4}. The experimental consequences depend on the surviving symmetry group and have been subject to numerous studies, see Refs.~\cite{K1,K2,Carroll,Jacobson,Mattingly,Altschul,Coleman}.

In this article we instead explore the consequences  of theories without an order parameter proposed in Ref.~\cite{CG}.  Studying the effects of VSR in terms of a Lagrangian field theory immediately runs into the obstacle that the operators that break Lorentz symmetry down to $H=Hom(2), Sim(2)$ are necessarily non-local in spacetime.    This means that the naive Feynman rules one would derive for such interactions, based on the underlying assumption that the quantum fields are causal, may not apply in VSR theories.   To circumvent this, we will consider processes (matrix elements) with at most one particle in the initial or final states.   Such observables are well described by ordinary quantum mechanics, and none of the off-shell ambiguities associated with non-local Feynman rules arise.

In order to construct such observables, we take as our starting the one-particle representations of the inhomogeneous group $ISim(2)$ ($Sim(2)$ plus spacetime translations).    We concentrate on $Sim(2)$ symmetry in which case the little group of massive particles consists of only the rotations about a preferred axis.  Thus massive particle representations are one-dimensional, labeled by spin about this axis and by the mass eigenvalue $m^2=P_\mu P^\mu$.  Consequently, states of different spins correspond to independent particles:   they can have different masses and interactions.

Given this, we then calculate the most general matrix elements of local operators (for instance the electromagnetic current $J^\mu(x)$) between one-particle states that are allowed by $Sim(2)$ symmetry.    We apply this result to study the effects of $Sim(2)$ symmetry on the electromagnetic properties of electrons.   Besides the usual electromagnetic form factors of $SO(3,1)$ invariant QED, several new terms consistent with VSR are allowed in the matrix elements of the electron current.    Lorentz invariant electrodynamics is recovered in the limit where these new form factors are set to zero.  The current matrix elements can then be used to derive an effective Hamiltonian for non-relativistic (NR) electrons interacting with electromagnetic fields.  Note however, that in the NR limit, the interactions allowed by VSR become local, and the construction of the Hamiltonian \emph{can} be achieved by the introduction of a three-vector spurion corresponding to a preferred spatial direction.   Thus the NR Hamiltonian consists of the most general set of electromagnetic moments that can be formed from this spurion together with the electron degrees of freedom.

Of the terms that arise in the NR theory, the strongest bounds are on the spin-dependent masses allowed by the $Sim(2)$ one-particle representation theory.   This effect manifests itself as a coupling of the form $H = \Delta m_e {\bf n}\cdot {\bf S}$.   This operator is constrained by torsion pendulum experiments, which place the bound $\Delta m_e/m_e < 10^{-26}$~\cite{HCCASS}.  The parameter $\Delta m_e$ can be set to zero by imposing $PT$ invariance on the theory.   However, $PT$ is not a symmetry of the Standard Model, and one expects that $PT$ violating effects are induced radiatively if not present in the theory from the beginning.   Naive estimates of the magnitude of such mass splittings exceed the experimental bounds and indicate that VSR theories need to be fine tuned.

\section{Symmetries}

Although the VSR groups do not admit invariant tensors, they select a preferred null direction which we take to be $n^\mu=(1,0,0,1)$.    There are four $SO(3,1)$ generators that have a simple action on $n$.   Defining ${\bar n}^\mu = (1,0,0,-1)$, $x^\mu=(0,1,0,0)$ and $y^\mu=(0,0,1,0)$, these generators can be written in the $(1/2,1/2)$ representation of $SO(3,1)$ as
\begin{eqnarray}
\label{eq:generators}
\nonumber
{T_1^\mu}_\nu &=& i (x^\mu n_\nu -n^\mu x_\nu)\\
\nonumber
{T_2^\mu}_\nu &=& i (y^\mu n_\nu -n^\mu y_\nu)\\
\nonumber
{J_3^\mu}_\nu  &=& i (x^\mu y_\nu - y^\mu x_\nu)\\
{K_3^\mu}_\nu &=&  {i\over 2} (n^\mu {\bar n}_\nu - {\bar n}^\mu n_\nu).
\end{eqnarray}
In an arbitrary representation, these generators satisfy the algebra $[T_1,T_2]=[J_3,K_3]=0$,  $[J_3,T_1]=iT_2,$  $[J_3,T_2]=-iT_1 $, $[K_3,T_1]=-iT_1$, $[K_3,T_2]=-iT_2$.   The group generated by $T_1$, $T_2$, and $J_3$ is the massless little group $E(2)$ which leaves $n$ invariant so is of no interest to us.   Including $K_3$ in the list of generators leads to groups with no invariant tensors.    The minimal group is $Hom(2)$, generated by $T_1$, $T_2$, $K_3$.   In addition, including $J_3$ leads to the group  $Sim(2)$.

Labeling the states in one irreducible representation of $SO(3,1)$ by its $J_3, K_3$ eigenvalues, the action of $T_1$ and $T_2$ gives a state that has $K_3$ lowered by one unit.   Therefore the only invariant subspaces under $T_{1,2}$ are either the whole representation, or the state with least $K_3$.    It follows that the irreducible representations of the VSR groups are one-dimensional, labeled by the $K_3$ eigenvalue.\footnote{In terms of the usual $(j_+,j_-)$ representations, the least $K_3$ state corresponds to $|m^{+}_z=j_+,m^{-}_z=-j_{-}\rangle$.}

We will build particle theories that are not invariant under the full Lorentz symmetry but only under $Hom(2)$ or $Sim(2)$.    As pointed out by Cohen and Glashow, this reduced set of spacetime symmetries is fully compatible with all kinematic tests of special relativity.   This follows simply from the fact that all inertial frames moving with $v<1$  can be reached by purely VSR transformations.   For example, the transformation
\begin{equation}
\label{eq:LV}
L(p)= \exp[-i\phi_1 T_1]\exp[-i\phi_2 T_2]\exp[-i\phi_3 K_3],
\end{equation}
with 
\begin{eqnarray}
 e^{-\phi_3}= {n\cdot p\over m}, &\phi_i = {p^i\over n\cdot p} (i=1,2),
\end{eqnarray}
takes a particle with momentum $p_0^\mu=(m,0,0,0)$ to one with momentum  $p^\mu=(E_{\bf p},{\bf p})$.  

The fact that any two  timelike vectors can be connected by a path in the VSR group makes it possible to construct massive one-particle representations of the inhomogeneous VSR groups $IHom(2)$ $ISim(2)$  ($Hom(2)$ and $Sim(2)$ extended by spacetime translation generators $P^\mu$) by the usual little group methods.    The little group for massive particles transfoming under $IHom(2)$ is the trivial group and the massive particle representations are one-dimensional.   For $ISim(2)$ the massive little group is the $U(1)$ subgroup generated by $J_3$.   Again the representations are one-dimensional, labeled by the spin eigenvalue $J_3$.   Thus the one-particle states are given by $|p,s\rangle$ with
\begin{eqnarray}
\nonumber
P^\mu |p,s\rangle = p^\mu |p,s\rangle,\\
J_3 |p,s\rangle = s |p,s\rangle.
\end{eqnarray}
A natural definition of states satisfying these properties is given by 
\begin{equation}
\label{eq:states}
|p,s\rangle = U(L(p)) |p_0,s\rangle,
\end{equation}
where $U(L(p))$ is the unitary operator corresponding to the boost in Eq.~(\ref{eq:LV}).    According to this definition, an $ISim(2)$ transformation $U(\Lambda)$ acts on the one-particle Hilbert space according to
\begin{equation}
U(\Lambda)  |p,s\rangle = \exp[-i s \theta(p,\Lambda)] |\Lambda p,s\rangle,
\end{equation}  
where the Wigner angle $\theta(p,\Lambda)$ is zero except for rotations about $z$.     

For example, in the framework of Ref.~\cite{CG}, the Hilbert space of  a single ``left-handed" massive neutrino is spanned by the states $|k,s=-1/2\rangle$.  The conventional states of spin-$1/2$ fermions, for instance the electron, correspond to different representations, and without introducing additional symmetries may be consistently assigned different values of $m^2=P_\mu P^\mu$,
\begin{equation}
P^2 |p,s=\pm1/2,e^-\rangle = m^2_{e\pm} |p,s=\pm1/2,e^-\rangle.
\end{equation}
One may note that the discrete symmetry 
\begin{equation}
S:  x^\mu\leftrightarrow y^\mu,
\end{equation}
takes $T_1\leftrightarrow T_2$, $K_3\rightarrow K_3$ and $J_3\rightarrow -J_3$, and is an automorphism of the $Sim(2)$ algebra.    This operation relates states of opposite $J_3$, and, if realized, would ensure that $m_{e+}=m_{e-}$.  However, this operation is equivalent to the discrete symmetry $PT$, which is broken explicitly in the Standard Model.   Thus if $Sim(2)$ is realized in nature, a generic feature is the splitting of masses within a spin multiplet.   The operation $S$ cannot be used to  relate masses within members of a larger $SO(3)$ multiplet.  (On the other hand, it is possible to obtain supersymmetric extensions of the  $ISim(2)$ algebra~\cite{susy1}.  This can be used to construct $ISim(2)$ theories in which fermionic and bosonic masses are related~\cite{susy1,susy2}.)

\section{Electromagnetic couplings}

In this section we explore the phenomenological consequences of  $Sim(2)$ symmetry for low-energy QED processes with electrons.   It is necessary to determine how the couplings of electrons to photons are modified by $Sim(2)$.  We will assume the existence of a local, conserved $Sim(2)$ 4-vector current operator $J^\mu(x)$ that couples to the electromagnetic field in the usual way.   To make contact with $SO(3,1)$ invariant electrodynamics, we parametrize the one-particle matrix elements of this current in terms of quantities transforming in the Dirac representation of $SO(3,1)$,      
\begin{equation}
\langle p' s' | J^\mu(x)|p s\rangle = e^{-i q\cdot x} {\bar u}_{s'}(p') \Gamma_{s's}^\mu(p',p) u_s(p),
\end{equation}
with $q=p-p'$, and $\Gamma^\mu_{s's}(p',p)$ a Dirac bilinear that we construct below.   If the one-particle states on the left-hand side of this equation are those of Eq.~(\ref{eq:states}) the spinors $u_s(p)$ must transform under $Sim(2)$ as~\cite{weinberg}
\begin{equation}
D(\Lambda) u_s(p) = e^{-i s\theta(p,\Lambda)} u_s(\Lambda p),
\end{equation} 
where $D(\Lambda)$ is the Dirac representation of the $Sim(2)$ transformation $\Lambda$.   This equation says in particular that the spinors at rest must be $J_z$ eigenstates
\begin{widetext}  
\begin{eqnarray}
u_{s=1/2}(p_0) = \sqrt{m_{e+}}\left(\begin{array}{c}
1\\
0\\
{m_+\over m_{e+}}\\
0
\end{array}\right), \ \ 
& u_{s=-1/2}(p_0)= \sqrt{m_{e-}}\left(\begin{array}{c}
0\\
{m_-\over m_{e-}}\\
0\\
1
\end{array}\right),
\end{eqnarray}
\end{widetext}
where the normalization is chosen for later convenience.   From these spinors we obtain the spinors in an arbitrary frame by boosting,
\begin{equation}
u_s(p) = D(L(p)) u_s(p_0).
\label{eq:spinors}
\end{equation}
This implies that the VSR spinors satisfy a $Sim(2)$ modified version of the Dirac equation~\cite{CG}
\begin{equation}
\left[\pslash - m_{\pm}  -\lambda_{\pm} {\nslash\over 2 n\cdot p}\right] u_{s=\pm 1/2}(p) =0.
\label{eq:VSR-Dirac}
\end{equation}
Here the parameters $\lambda_s$ are related to the physical masses by $m^2_{e\pm} = m_\pm^2+\lambda_\pm$.    The non-local nature of the Lorentz violating terms is due to the non-existence of $Sim(2)$ invariant tensors, as pointed out in~\cite{CG}.

The matrices $\Gamma^\mu_{s's}(p',p)$ are determined by $Sim(2)$ invariance and current conservation, $q_\mu \Gamma^\mu_{s's}(p',p)=0$.  Keeping only Dirac bilinears that give rise to terms at most linear in $q^\mu$ one finds
\begin{widetext}
\begin{eqnarray}
\nonumber
{\bar u}_{s'}(p') \Gamma^\mu_{s's}(p',p) u_{s}(p) &=& 
{\bar u}_{s'}(p')\left(\gamma^\mu  + \lambda_s \delta_{ss'} {\nslash n^\mu\over 2 (n\cdot p) (n\cdot p')}\right) u_{s}(p) \\
\nonumber
& &  {}+  q_s\delta_{s's}{\bar u}_{s'}(p') \gamma_5 \left(\gamma^\mu +\lambda_s {\nslash n^\mu\over 2 (n\cdot p) (n\cdot p')} +  {2 m\over n\cdot (p+p')} i\sigma^{\mu\nu} n_\nu\right)u_{s}(p) \\
\nonumber
& &  {}+{i\over 2m}  {\bar u}_{s'}(p') (F^{s's}_2 +  i \gamma_5 F^{s's}_3) \sigma^{\mu\nu} q_\nu  u_{s}(p)\\
\nonumber
& &  {}+  {im F^{s's}_5\over (n\cdot p)(n\cdot p')} \left( n\cdot q \sigma^{\mu\nu} n_\nu - n^\mu \sigma^{\nu \rho} q_\nu n_\rho \right) u_{s}(p) \\
\nonumber
& & {} + {im F^{s's}_6\over (n\cdot p)(n\cdot p')} {\bar u}_{s'}(p')\gamma_5 \left( n\cdot (p+p') n^\mu + i n\cdot q\sigma^{\mu\nu} n_\nu\right)u_{s}(p) \\
\nonumber 
& & {} + {F^{s's}_7\over n\cdot (p+p')} {\bar u}_{s'}(p')\gamma_5\left( n\cdot q \gamma^\mu - 2 m n^\mu  +\lambda_s {\nslash n^\mu\over 2 (n\cdot p) (n\cdot p')} n\cdot q\right) u_{s}(p)\\
& &  {} +  {\bar u}_{s'}(p') i (F^{s's}_8 -  \gamma_5 F^{s's}_9) {n\cdot (p+p')\over (n\cdot p) (n\cdot p')} \epsilon^\mu_{\nu\alpha\beta} q^\nu n^\alpha \gamma^\beta u_{s}(p).
\label{eq:current}
\end{eqnarray}
\end{widetext}
Note that in this equation, we have tuned the Dirac masses $m_+=m_-=m$. This is required by  the Ward identity.  The second line of this equation contain terms that survive in the $q^\mu\rightarrow 0$ limit, and corresponds to spin-dependent electromagnetic charges for the electron.    Although this is certainly consistent with the one-particle representations of $ISim(2)$, it excludes spin-flipping electromagnetic transitions.   Consequently, the fine-tuning $q_s=0$ is required to obtain an acceptable phenomenology. In addition, this term may be inconsistent quantum mechanically due to anomalies.

 In order to find the non-relativistic (NR) Hamiltonian for electrons interacting with background electromagnetic fields, we match the $q^\mu\rightarrow 0$ limit of the scattering amplitude
\begin{equation}
\langle p',s'| iT|p,s\rangle  = ie \int d^4 x e^{-iq\cdot x} A_\mu(x) {\bar u}_{s'}(p') \Gamma_{s's}^\mu(p',p) u_s(p)
\end{equation}
to the same result in the NR theory.   Including the NR expasion of the particle energies as well, the part of the NR Hamiltonian that violates Lorentz symmetry is $H_{VSR}=H_{\Delta m}+H_{EM}$, where
\begin{eqnarray}
\label{eq:deltam}
H_{\Delta m} =   \Delta m_e {\bf n}\cdot {\bf S},
\end{eqnarray}
and $H_{EM}$ contains, in addition to the conventional electromagnetic couplings, operators of the form
\begin{widetext}
\begin{eqnarray}
\nonumber
H_{EM} &=& - {e\Delta m_e\over 2 {\bar m}^2_e} ({\bf n}\cdot {\bf S})  \left({\bf p}\cdot {\bf A} +{\bf A}\cdot {\bf p}\right) -  {e\over 2 {\bar m}_e} {\bf n}\cdot (g_{B_\parallel} {\bf B} +  g_{E_\parallel} {\bf E}) +  {e\over 2 {\bar m}_e} {\bf S}\cdot  (g_{E_\perp}  {\bf n}\times {\bf E} + g_{B_\perp}  {\bf n}\times {\bf B})\\
 & & {} +  {e\over 2 {\bar m}_e} ({\bf n}\cdot {\bf S}) \left(g'_{B_\parallel} ({\bf n}\cdot {\bf B}) + g'_{E_\parallel} ({\bf n}\cdot {\bf E})\right).
 \label{eq:HNR}
\end{eqnarray}
\end{widetext}
Here ${\bar m}_e=(m_{e+}+m_{e-})/2$, ${\Delta m}_e=m_{e+}-m_{e-}$.   The coupling constants $g_{B_\parallel,E_\parallel}$, $g_{B_\perp,E_\perp}$, and $g'_{B_\parallel,E_\parallel}$ can be expressed as linear combinations of the VSR parameters $\lambda_{\pm}$, and $F^{s's}_i$ in a way that is calculable from Eq.~(\ref{eq:current}).   These expressions are not needed in what follows\footnote{In order to cast $H_{EM}$ in a form that has a convetional limit as the Lorentz breaking parameters go to zero, it is necessary to perform unitary transformation on the spin operator acting on $Sim(2)$ states.    See the appendix for details.}.   

Note that the electromagnetic couplings generated by the $Sim(2)$ covariant current $J^\mu(x)$ comprise the most general set of NR operators that can be constructed from the electromagnetic field and the `spurion' represented by the preferred direction ${\bf n}$.   This follows from the observation that  in the NR limit, the non-local terms proportional to powers of $1/(n\cdot p)$ have an expansion which is analytic about zero three-momentum and thus correspond to local NR operators.  Therefore, the structure of  the low energy $Sim(2)$ QED Hamiltonian is largely indistinguishable from that of theories in which Lorentz symmetry is spontaneously broken.

Nevertheless, it is possible that $Sim(2)$ symmetry implies correlations among the low energy couplings appearing in the Hamiltonian $H_{EM}$.   For example, it is argued in Ref.~\cite{CG} that $SU(2)_L$ gauge invariance may relate VSR neutrino masses to the parameters $\lambda_{\pm}$ in the VSR-modified Dirac equation for the electron.   If all other electromagnetic form factors in Eq.~(\ref{eq:current}) were set to zero, this would yield a value  ${g}_{E\perp}\sim 10^{-15}$ near current  experimental limits.   The question of whether  ${g}_{E\perp}$ arises solely from neutrino mass effects or partially from other $Sim(2)$ effects (encoded in the coefficients $F_i$) cannot be addressed without knowing more details about the specific mechanism that would generate $Sim(2)$ rather than $SO(3,1)$ as the spacetime symmetry of nature.

Of the parameters appearing in the NR Hamiltonian, Eqs.~(\ref{eq:deltam}), (\ref{eq:HNR}), the most stringent bound is on the parameter $\Delta m_e$.   This comes from a recent torsion pendulum experiment with a macroscopic number of polarized electrons~\cite{HCCASS}.   This experiment looks for torques induced by the Earth's motion relative to the frame in which the direction ${\bf n}$ is constant.  Failure to detect such a signal puts a bound $\Delta m_e/m_e < 10^{-26}$.   We expect the limits on the
electric moments to be comparable to the limit on the ordinary electric moment that is $g_{E}<10^{-15}$~\cite{demille}.  It would be interesting to investigate what the experimental bounds on the magnetic moments are, but we do not expect those bounds to be as tight as the bounds on the electric moments.


\section{Conclusion}

 If nature is described by $Sim(2)$ theories, massive particle multiplets are one-dimensional, labeled by spin along a preferred axis, and particle properties become spin dependent.   In this paper, we have explored the consequences of this reduced symmetry on the low energy properties of electrons.   We find that experiments put strong bounds on possible symmetry breaking effects.   In particular, the $PT$ violating spin dependent mass difference is bounded to be $\Delta m_e/m_e <10^{-26}$.  Such effects would be generated radiatively by Standard Model loops and are likely to exceed this bound, indicating that some fine tuning may be necessary.

In this paper we have focused on implications of $Sim(2)$ symmetry at low energies, where calculations based on one-particle quantum mechanics are sufficient.  In order to understand the full consequences of VSR, however, several theoretical issues remain.   For instance,  it is not clear whether a multi-particle   quantum theory based on one-particle representations of $ISim(2)$ (or equivalently, second quantization of the non-local Dirac equation) can be constructed in terms of causal quantum fields.   Even if such a field theoretic description is found, it would be non-trivial to formulate it in a way compatible with general coordinate invariance.   Thus the question of whether $Sim(2)$ symmetry can be made consistent with the Standard Model plus general relativity is still an open one.

\section{Acknowledgements}

We thank A. Cohen  and  D. DeMille for helpful discussions.    
This work was supported in part by the Department of Energy under Grant DE-FG-02-92ER40704.  
 W.G. and W.S. thank the Aspen Center for Physics where some of this work was conducted.

{\bf Note added}:   While this paper was being completed, Ref.~\cite{mehen}, which has some overlap with the work presented here, appeared on the arXiv.

\begin{appendix}
\section{Spin redefinitions}

In taking the naive low energy limit of Eq.~(\ref{eq:current}) one ecounters the following term 
\begin{equation}
H_{E_\perp} = - {e\over 2 {\bar m}_e}\sigma\cdot ({\bf n}\times {\bf E}),
\end{equation}
which arises from the $\gamma^\mu$ part of the matrix element.   This would seem to be in stark contrast with experiment.   However, in  the Lorentz limit, it is possible to define a spin operator acting on the states of Eq.~(\ref{eq:states}) by
\begin{equation}
{S}^i_{s's} = {1\over 2 E_{\bf p}} u^\dagger_{s'}(p) 
\left(\begin{array}{cc}
{\sigma}^i/2 & 0\\
0 & {\sigma}^i/2
\end{array}\right) u_s(p),
\end{equation}
which satisfies the $SU(2)$ commutation relations.   Using the spinors corresponding to $Sim(2)$ one-particle states, one finds that to linear order in momentum, this operator has the following form on NR states
\begin{equation}
\label{eq:usspin}
{\bf S} = {1\over 2}{\bf \sigma} + {1\over 2 m} {\bf \sigma} \times ({\bf n}\times {\bf p}) + {\cal O}({\bf p}^2).
\end{equation}
In order to identify the terms in $H_{NR}$ with electromagnetic moments as quoted by precision atomic physics experiments, it is necessary to use a conventionally defined spin operator ${\bf S}_c=\sigma/2$.   To linear order in $1/m$, such an operator is related to the spin operator of Eq.~(\ref{eq:usspin}) by a unitary transformation ${\bf S}_c = U {\bf S} U^{\dagger},$ where
\begin{equation}
U = \exp\left[ {i\over m}\sigma\cdot ({\bf n}\times {\bf p})\right].
\end{equation}
(Note that in the presence of a background field, the correct unitary transformation requires the substitution ${\bf p}\rightarrow {\bf p} + e {\bf A}(x)$.).  In this new frame, states evolve according to a new Hamiltonian
\begin{eqnarray}
\nonumber
H_c &=&  U H U^{\dagger} + i {\dot U} U^\dagger\\
        &=& H + {e\over 2 m}\sigma\cdot ({\bf n}\times {\bf E}) + {\cal O}(1/m^2).
\end{eqnarray}
Thus in the frame in which the spin is conventional, the effects of the operator $H_{E_\perp}$, are cancelled to zeroth order in the Lorentz breaking terms.  

\end{appendix}

\end{document}